\documentclass[12pt]{article}
\usepackage{authblk}
\usepackage[bookmarksnumbered, colorlinks, plainpages]{hyperref}
\usepackage{amsmath, amsthm, amscd, amsfonts, amssymb, graphicx, color, booktabs}
\newcommand{\Bzsmm}{$\mathrm{B^0_{(s)}}\to \mu^+\mu^-$}
\newcommand{\Bsmm}{$\mathrm{B^0_s}\to \mu^+\mu^-$}
\newcommand{\Bdmm}{$\mathrm{B^0}\to \mu^+\mu^-$}
\newcommand{\Bs}{$\mathrm{B^0_s}$}

\usepackage{hyperref}

\numberwithin{equation}{section}
\definecolor{email}{rgb}{0.00,0.00,0.84}
\begin{document}
\setcounter{page}{1}

\title{\large \bf 12th Workshop on the CKM Unitarity Triangle\\ Santiago de Compostela, 18-22 September 2023 \\ \vspace{0.3cm}
\LARGE \Bzsmm\  rare decays at LHC}

\author{Giacomo Fedi\textsuperscript{1} on behalf of the ATLAS, CMS, and LHCb Collaborations \\
    \texttt{giacomo.fedi@cern.ch} \\
        \textsuperscript{1}Imperial College London, United Kingdom \\ }

\maketitle

\begin{abstract}
The ATLAS, CMS, and LHCb collaborations have leveraged data from proton-proton collisions at the Large Hadron Collider (LHC) to advance our understanding of fundamental particles and their interactions. These collaborative efforts underscore the significance of integrating datasets across different runs and energy levels, yielding insights that contribute to the ongoing measurements of the \Bzsmm{} decay branching fractions and \Bsmm{} effective lifetime.

\end{abstract} \maketitle

\section{Introduction}

In the Standard Model (SM), the \Bzsmm{} decays are governed by flavour-changing neutral current (FCNC) transitions and are further hindered by helicity suppression. The \Bdmm{} decay, in particular, faces additional suppression relative due to the factor $|V_{\mathrm{tb}}/V_{\mathrm{ts}}|^2$, approximately 20 times lower branching fraction than the \Bsmm{} decay. In both transitions manifestation of New Physics (NP) could occur in the loop diagrams enhancing or reducing the branching fractions (BFs).

The search for NP contributions is then achieved through precise measurements of BFs and the lifetime of \Bs{} mesons. Moreover the search for NP is facilitated by precise SM expectations of the \Bzsmm{} BFs~\cite{Beneke:2019slt} :

\begin{equation*}
    \mathcal{B(\mathrm{B^0_s}\to \mu^+\mu^-)}=(3.66\pm 0.14)\times 10^{-9},
\end{equation*}
\begin{equation*}
    \mathcal{B(\mathrm{B^0}\to \mu^+\mu^-)}=(1.03\pm 0.05)\times 10^{-10}.
\end{equation*}

The heavy and light \Bs{} mass eigenstates are linear combinations of the flavor eigenstates. The lifetimes of the heavy and light mass eigenstates are $\tau_{\mathrm{H}} =
1.624\pm0.009$ ps and $\tau_{\mathrm{L}} = 1.429 \pm0.007$ ps, respectively~\cite{ParticleDataGroup:2022pth}. In absence of CP violation in the mixing, the \Bsmm{} decay is only accessible by the \Bs{} heavy mass eigenstate, as it  coincides with the CP-odd eigenstate. Thus, measuring an effective lifetime of the \Bsmm{} which is not in accordance with a pure heavy mass eigenstate lifetime would be a hint of NP.

\begin{figure} [h!]
  \begin{center}
    \includegraphics[width=420pt]{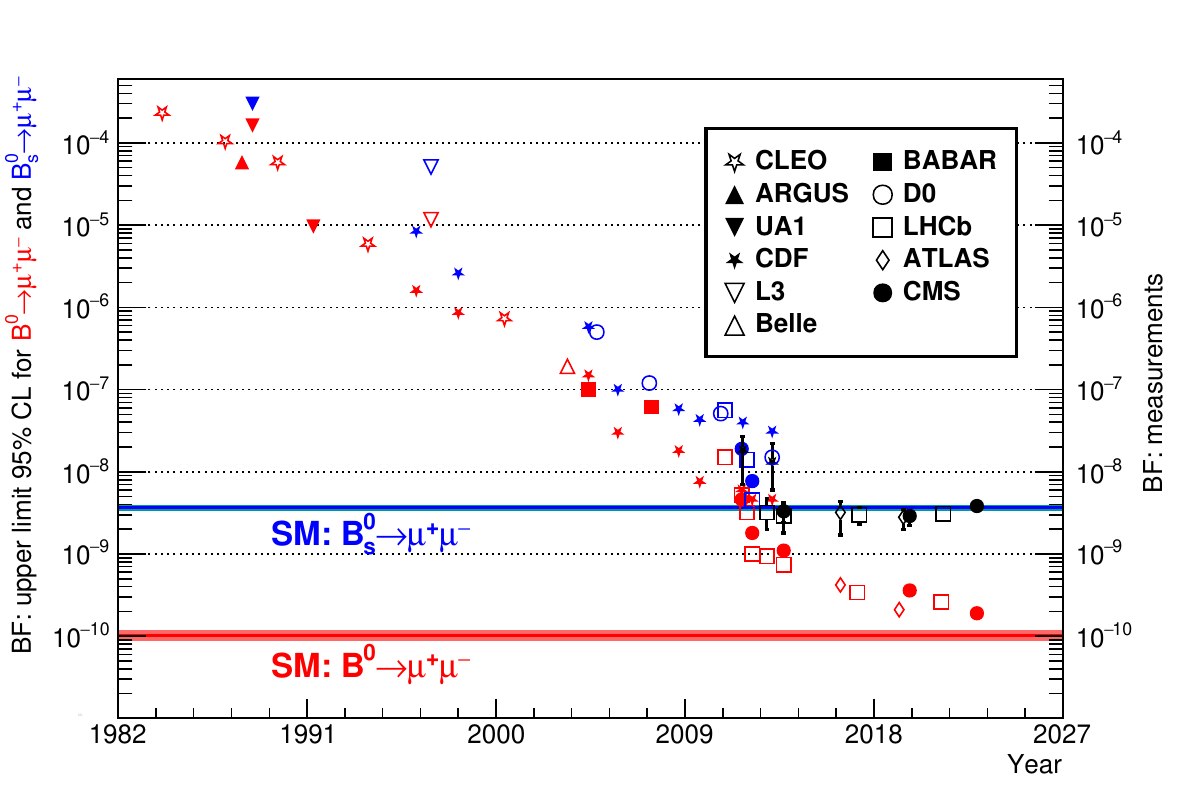}
    \caption{Evolution with time of the 95\% CL upper limit on \Bdmm{} (red) and \Bsmm{} (blue) branching fraction. \Bsmm{} branching fraction measurements (black). Results by several experiments~\cite{CMS:2022mgd,LHCb:2021vsc,ATLAS:2018cur,CMS:BPH}. }
    \label{fig:evolution}
  \end{center}
\end{figure}

\section{Branching fraction measurements}

The ATLAS~\cite{ATLAS:2008xda}, CMS~\cite{CMS:2008xjf}, and LHCb~\cite{LHCb:2008vvz} collaborations have leveraged data from proton-proton collisions to measure the \Bzsmm{} BFs. ATLAS utilised proton-proton collision data from 2015 to 2016 (partial Run2), amounting to 26.3 fb$^{-1}$ at a centre-of-mass energy of 13 TeV, which was combined with Run1 results at the likelihood level~\cite{ATLAS:2018cur}. CMS employed data from 2016 to 2018 (Run2), consisting of 140 fb$^{-1}$ at 13 TeV, excluding previous data due to minimal benefits~\cite{CMS:2022mgd}. LHCb combined data from 2011-2012 (Run1) and 2015-2018 (Run2), for a total of 9 fb$^{-1}$ at various energies (7, 8, and 13 TeV), and integrated with previous analyses at the likelihood level~\cite{LHCb:2021vsc}.

The \Bzsmm{} branching fractions are measured relative to a normalisation channel. Most common normalisation channel in the different analyses is the $\mathrm{B^+\rightarrow\mathrm{J}/\psi\mathrm{K^+}}$ channel, although other normalisation channels are also used:
\begin{equation*}
\mathcal{B}(\mathrm{B^0_s}\to \mu^+\mu^-)=\frac{n^{obs}_{\mathrm{B^0_s}}}{\epsilon\mathcal{L}\sigma(pp\to \mathrm{B^0_s})}=\frac{n^{obs}_{\mathrm{B^0_s} }}{n^{obs}_{\mathrm{B^+}}}
\frac{\epsilon_{\mathrm{B^+\rightarrow\mathrm{J}/\psi\mathrm{K^+}}}}{\epsilon_{\mathrm{B^0_s}\rightarrow\mu^+\mu^-}}
 \frac{f_\mathrm{u}}{f_\mathrm{s}}\mathcal{B}(\mathrm{B^{+}}\to \mathrm{J}/\psi\mathrm{K^{+}}),
\end{equation*}

\begin{equation*}
\mathcal{B}(\mathrm{B^0}\to \mu^+\mu^-)=\frac{n^{obs}_{\mathrm{B^0}}}{\epsilon\mathcal{L}\sigma(pp\to \mathrm{B^0})}=\frac{n^{obs}_{\mathrm{B^0} }}{n^{obs}_{\mathrm{B^+}}}
\frac{\epsilon_{\mathrm{B^+\rightarrow\mathrm{J}/\psi\mathrm{K^+}}}}{\epsilon_{\mathrm{B^0}\rightarrow\mu^+\mu^-}}
\frac{f_\mathrm{u}}{f_\mathrm{d}}\mathcal{B}(\mathrm{B^{+}}\to \mathrm{J}/\psi\mathrm{K^{+}}),
\end{equation*}
where $\epsilon$ is the total efficiency, $\mathcal{L}$ is the integrated luminosity, $\sigma(z)$ is the cross section of the process $z$, $f_x$ is the fragmentation factor of the quark $x$, $n^{obs}_{\mathrm{B}}$ is the number of observed meson $\mathrm{B}$, and $\epsilon_y$ is the efficiency measured in channel $y$.  

All the analyses select the events to reduce the background contribution and increase the signal purity. Machine learning techniques - boost decision trees - are used to improve the selection and are also used to categorise the events to improve the fit sensitivity. To obtain the observed number of mesons a maximum likelihood fit on data is applied. The likelihood function is composed of a signal model and a number of background models (partial decays, peaking, and combinatorial).

The systematic uncertainties are evaluated according to the assumptions and techniques implemented in each analysis and their relative value is generally below 10\%, anyway lower than the statistical uncertainty. A common systematic uncertainty, which is also one of the biggest in all the measurements (3-5\%), stems from the assumptions made on the $f_{\mathrm{s}}/f_{\mathrm{u}}$ fragmentation ratio. This ratio is obtained from the literature~\cite{HFLAV:2016hnz,LHCb:2021qbv} and at the moment is considered a source of an irreducible uncertainty. 

\begin{figure} [h!]
  \begin{center}
    \includegraphics[width=420pt]{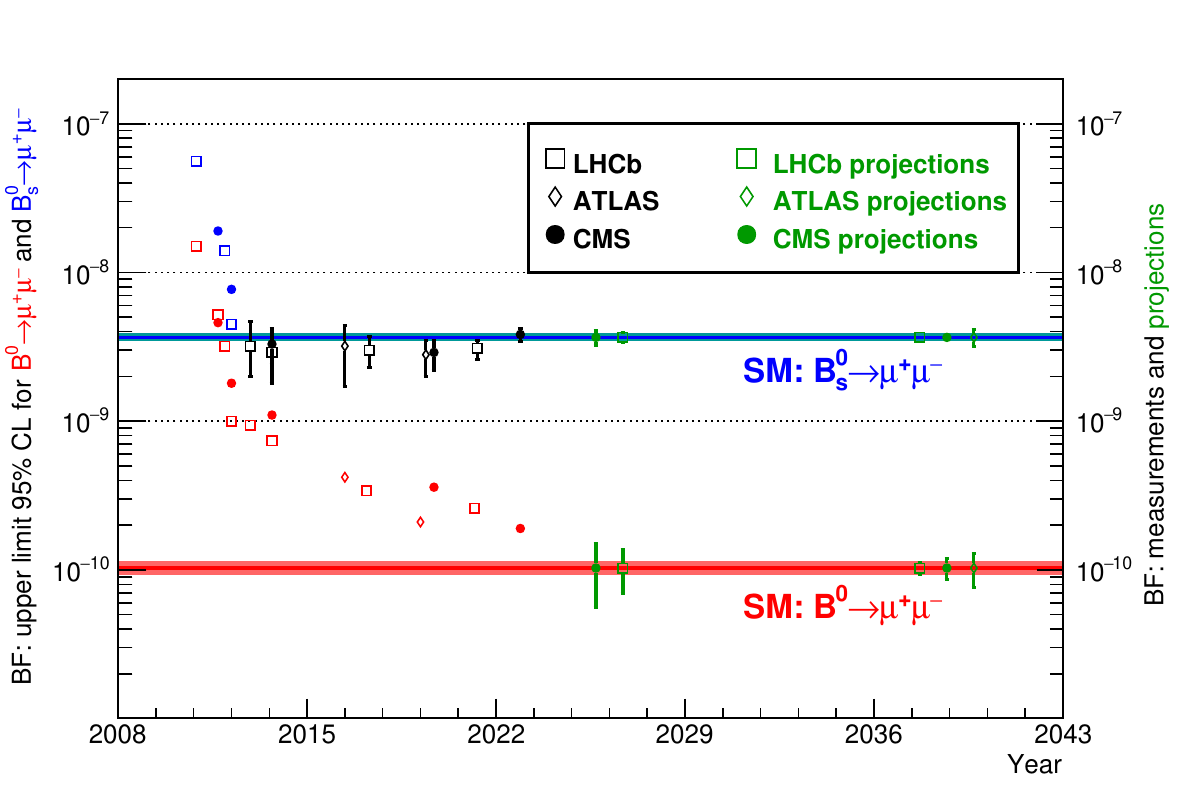}
    \caption{Evolution with time of the 95\% CL upper limit on \Bdmm{} (red) and \Bsmm{} (blue) branching fraction. \Bsmm{} branching fraction measurements (black). In green, projections for future. CMS, ATLAS and LHCb results~\cite{CMS:2022mgd,LHCb:2021vsc,ATLAS:2018cur,CidVidal:2018eel,CMS:BPH}.}
    \label{fig:future}
  \end{center}
\end{figure}

The Run2 collision data for a total of 26 fb$^{-1}$ was combined by ATLAS with the Run1 BFs measurements:
\begin{equation*}
    \mathcal{B(\mathrm{B^0_s}\to \mu^+\mu^-)}=(2.8^{+0.8}_{-0.7})\times 10^{-9},
\end{equation*}
\begin{equation*}
    \mathcal{B(\mathrm{B^0}\to \mu^+\mu^-)}<2.1\times 10^{-10}\ @ 95\% CL,
\end{equation*}
both results are compatible with the SM expectation withing 2.4$\sigma$.

The Run1 and Run2 collision data for a total of 9 fb$^{-1}$ were analysed by LHCb to measure the BFs:
\begin{equation*}
    \mathcal{B(\mathrm{B^0_s}\to \mu^+\mu^-)}=\left[3.09^{+0.46}_{-0.43}(stat)^{+0.15}_{-0.1=1}(syst)\right]\times 10^{-9},
\end{equation*}
\begin{equation*}
    \mathcal{B(\mathrm{B^0}\to \mu^+\mu^-)}< 2.6\times 10^{-10}\ @ 95\% CL,
\end{equation*}
\begin{equation*}
    \mathcal{B(\mathrm{B^0_s}\to \mu^+\mu^-\gamma)} < 2.0\times 10^{-9}\ @ 95\% CL,
\end{equation*}
all measurements are compatible with SM. Notably, LHCb also measured the $\mathrm{B^0_s}\to \mu^+\mu^-\gamma$ branching fraction in the $\mu\mu$ mass range $<4.9$ GeV.

Run2 collision data for a total of 140 fb$^{-1}$ were analysed by CMS to measure the BFs:
\begin{equation*}
    \mathcal{B(\mathrm{B^0_s}\to \mu^+\mu^-)}=\left[3.83^{+0.38}_{-0.36}(stat)^{+0.19}_{-0.16}(syst)^{+0.14}_{-0.13}(f_\mathrm{s}/f_\mathrm{u})\right]\times 10^{-9},
\end{equation*}
\begin{equation*}
    \mathcal{B(\mathrm{B^0}\to \mu^+\mu^-)}< 1.9\times 10^{-10}\ @ 95\% CL,
\end{equation*}
all measurements are in agreement with the SM. To date, these are are most precise single measurements in the literature. 

In fig.~\ref{fig:evolution} the time evolution of the branching fraction measurements by different experiments is shown. The \Bsmm{} branching fraction has been observed the first time by CMS and LHCb combining their data sets in 2014~\cite{CMS:2014xfa}. In fig.~\ref{fig:future} the time evolution of the branching fraction measurements made at the LHC experiments with the extrapolations made for Run3 and HL-LHC data is shown.

\section{Lifetime measurements}

The three experiments measured the lifetime on the same data sets used for the BFs measurements, although some additional selections are made to reduce the background contributions. The maximum likelihood function is thus simplified and fits both the mass and the lifetime distributions. 

LHCb measured a effective lifetime of the \Bs{} meson decaying in \Bsmm{} compatible within 1.5$\sigma$ with the pure CP-odd hypothesis and compatible within 2.2$\sigma$ with the pure CP-even hypothesis:
\begin{equation*}
             \tau = 2.07\pm 0.29(stat)\pm 0.03(syst)\ ps.
\end{equation*}
Also CMS measured the effective lifetime compatible with SM expectation:
\begin{equation*}
         \tau = 1.83^{+0.23}_{-0.20}(stat)^{+0.04}_{-0.04}(syst)\ ps.
\end{equation*}
ATLAS implemented a different fit technique~\cite{ATLAS:2023trk}. The $\mu\mu$ invariant mass distribution is fitted by using the S-plot technique and the \Bs{} lifetime background-subtracted distribution is fitted by using simulation-generated templates. The template that minimises the $\chi^2$ is considered the best fit and the effective lifetime value used for its generation to be the measured \Bs{} lifetime value:
\begin{equation*}
         \tau = 0.99^{+0.42}_{-0.07}(stat)\pm 0.17(syst)\ ps,
\end{equation*}
which is in agreement with the SM expectation.

\section{Summary}
ATLAS, CMS, and LHCb collaborations measured the BFs of the \Bzsmm{} decays and the effective lifetime of the \Bs{} in the \Bsmm{} decay channel. The current best measurement of the branching fraction has been published by CMS. All the experiments set upper limits at 95\% CL on the \Bdmm{} branching fraction. We expect to observe a statistically significant \Bdmm{} branching fraction for the first time once the Run3 data will be analysed~\cite{CidVidal:2018eel}. The \Bs{} effective lifetime has large statistical uncertainty and to exclude the presence of a \Bs{} light mass eigenstate contributions in the \Bsmm{} decay, more statistics is needed, possibly by the end of HL-LHC.


\bibliographystyle{unsrt}

\begin{thebibliography}{99}


\bibitem{Beneke:2019slt}
M.~Beneke, C.~Bobeth and R.~Szafron,
JHEP \textbf{10} (2019), 232
[erratum: JHEP \textbf{11} (2022), 099]
doi:10.1007/JHEP10(2019)232
[arXiv:1908.07011 [hep-ph]].

\bibitem{CMS:2008xjf}
S.~Chatrchyan \textit{et al.} [CMS],
JINST \textbf{3} (2008), S08004
doi:10.1088/1748-0221/3/08/S08004

\bibitem{ATLAS:2008xda}
G.~Aad \textit{et al.} [ATLAS],
JINST \textbf{3} (2008), S08003
doi:10.1088/1748-0221/3/08/S08003

\bibitem{LHCb:2008vvz}
A.~A.~Alves, Jr. \textit{et al.} [LHCb],
JINST \textbf{3} (2008), S08005
doi:10.1088/1748-0221/3/08/S08005

\bibitem{ParticleDataGroup:2022pth}
R.~L.~Workman \textit{et al.} [Particle Data Group],
PTEP \textbf{2022} (2022), 083C01
doi:10.1093/ptep/ptac097

\bibitem{CMS:2022mgd}
A.~Tumasyan \textit{et al.} [CMS],
Phys. Lett. B \textbf{842} (2023), 137955
doi:10.1016/j.physletb.2023.137955
[arXiv:2212.10311 [hep-ex]].

\bibitem{LHCb:2021vsc}
R.~Aaij \textit{et al.} [LHCb],
Phys. Rev. Lett. \textbf{128} (2022) no.4, 041801
doi:10.1103/PhysRevLett.128.041801
[arXiv:2108.09284 [hep-ex]].

\bibitem{ATLAS:2018cur}
M.~Aaboud \textit{et al.} [ATLAS],
JHEP \textbf{04} (2019), 098
doi:10.1007/JHEP04(2019)098
[arXiv:1812.03017 [hep-ex]].

\bibitem{HFLAV:2016hnz}
Y.~Amhis \textit{et al.} [HFLAV],
Eur. Phys. J. C \textbf{77} (2017) no.12, 895
doi:10.1140/epjc/s10052-017-5058-4
[arXiv:1612.07233 [hep-ex]].

\bibitem{LHCb:2021qbv}
R.~Aaij \textit{et al.} [LHCb],
Phys. Rev. D \textbf{104} (2021) no.3, 032005
doi:10.1103/PhysRevD.104.032005
[arXiv:2103.06810 [hep-ex]].

\bibitem{CMS:2014xfa}
V.~Khachatryan \textit{et al.} [CMS and LHCb],
Nature \textbf{522} (2015), 68-72
doi:10.1038/nature14474
[arXiv:1411.4413 [hep-ex]].

\bibitem{ATLAS:2023trk}
G.~Aad \textit{et al.} [ATLAS],
JHEP \textbf{09} (2023), 199
doi:10.1007/JHEP09(2023)199
[arXiv:2308.01171 [hep-ex]].

\bibitem{CidVidal:2018eel}
X.~Cid Vidal, M.~D'Onofrio, P.~J.~Fox, R.~Torre, K.~A.~Ulmer, A.~Aboubrahim, A.~Albert, J.~Alimena, B.~C.~Allanach and C.~Alpigiani, \textit{et al.}
CERN Yellow Rep. Monogr. \textbf{7} (2019), 585-865
doi:10.23731/CYRM-2019-007.585
[arXiv:1812.07831 [hep-ph]].

\bibitem{CMS:BPH}
https://twiki.cern.ch/twiki/bin/view/CMSPublic/PhysicsResultsBPH

\end{thebibliography}

\end{document}